# A dynamic analysis of magnetized plasma sheath in a collisionless scenario with ion sources

S. Adhikari [a)], R. Moulick, K. S. Goswami

*Centre of Plasma Physics – Institute for Plasma Research, Nazirakhat, Sonapur, Assam, India – 782402*

## ABSTRACT

In this paper, the central concern is to analyze the influence of the forces that controls the ion dynamics inside a magnetized plasma sheath under collisionless conditions. The ionization effects are considered within the sheath. The magnetic field is tilted in x-z plane and makes an angle with the x-axis. The motivation of the paper is to see the effect of both field strength variation and the variation of the inclination angle on the force fields inside the sheath. The pitch length and pitch angle for the particle velocity field is also calculated and has been found to vary widely with the inclination angle and the strength of the magnetic field. The role of the Lorentz force and energy acquired by the ions while moving towards the wall is highlighted. A comparison between two different ion sources has also been foregrounded.

___________________________________

a) Email: sayanadhikari207@gmail.com

I.  INTRODUCTION

The study of plasma sheath is an old but important issue in plasma physics. Introduction of the magnetic field makes the problem more complex and interesting. Magnetized plasma sheath has been studied extensively in plenty of works[1-8]. The study of particle dynamics in plasma sheath for the inclined magnetic field was first attempted by Chodura[7] in 1982. He pointed out the fact that the potential distribution in the presence of the inclined magnetic field consists of two different scaled structure. First, the usual electrostatic sheath, known as the Debye sheath and the second is the magnetic pre-sheath also known as Chodura sheath. The Chodura sheath is of the order of ion gyroradius. In a scenario, where the field lines intersect the wall surface making some angle, the strong electric field present in the CS as well as in DS turns the ion flow from parallel magnetic field direction to the direction perpendicular to the wall. Hence, the study of forces influencing the ion dynamics in the sheath is a requirement to understand the sheath structure. There are several issues of sheath physics like in particular, the modulation of the sheath to the perturbation of different plasma parameters in the core plasma or the instabilities caused by the deviations of distribution functions in the sheath from a Maxwellian[2,7] can only be explained studying the governing forces that manipulate the ion dynamics. This kind study of the magnetized sheath is not a new subject. There are several authors who have presented their views on the same topic with some excellent inputs. The motivation of this work came during the analysis of few related work worth mentioning, [9-14]. All of this paper addressed almost all the issues that can arise from the work.  The things that need to be discussed are the dynamic behavior of the force fields inside the sheath and contribution of the ion source term. Most of the papers have taken the assumption that ions are isothermal and electrons are Boltzmann distributed. This implies that there should not be any creation or destruction of ions too as the electrons follow Boltzmann relation. But all of these work did not take this into account and dealt the problem with zero source term. Hence these models are partially inconsistent. This fact is also supported by the Gyergyek et al[15]. Two types of source term have been used in this work and a comparison of their contributions has been demonstrated. Need to mention that, this work does not include the effect of the ion-neutral collision. Although collision is a dominant process in such systems, the decision was necessary to omit collision in the work as it would shield the main concern of the paper.

As we have already mentioned that the problem aged a lot. There are some benefits one can get out of this. There are lots of literature available related to it. After a literature review, it can be summarized as follows. In the same work in 1982, Chodura[7] established the fact that the total potential drop is independent of the angle of inclination ($\alpha$). In 1993 Holland et al.[16] worked in a similar plasma environment keeping $\alpha$ within the range $0° \leq \alpha \leq 9°$ and system length ($L_D$) within the range $r_{ci} \leq L_D \leq r_{ce}$, where $r_{ci}$ represents ion cyclotron radius and $r_{ce}$ represents electron cyclotron radius. They observed the transition of wall potential from positive to negative in the presence of $\mathbf{E} \times \mathbf{B}$ drift in the bulk plasma towards the wall for very small inclination angle. The angle of inclination can be classified into two categories – oblique incidence and grazing incidence. In the case of an oblique incidence, the angle ($\alpha$) between the magnetic field and the surface is larger than $10°$ and the wall potential which has been considered as floating is weakly dependent on $\alpha$, where in the case of grazing incidence it is in the range of $0° \leq \alpha \leq 10°$ and strongly dependent on $\alpha$. In 1993 Bergmann[17] did a 2D-3V particle in cell simulation of a flat Langmuir probe mounted into a particle absorbing plate to study the influence of an obliquely oriented strong magnetic field i.e. the relation between Larmor radii and Debye length. Their observation also supports the two region sheath theory reported by others. In 1998 Cohen and Ryutov[18] provided an assortment of particle orbits in a sheath in the presence of an inclined magnetic field. They established a general solution for the ion distribution function at random point of the sheath. The possible role of an electrostatic field directed along the wall was also mentioned.

At this point, it is important to mention about the two different approaches used by the different group of people to study the problem of the magnetized sheath. One approach is shown by Riemann[1] in 1994 where all the components of momentum equation are coupled to get a single equation in an integral form for the collisionless case. Another approach is to solve the momentum equation by solving simultaneously with Poisson's equation and continuity equation. Despite the fact that the latter approach has some constraints regarding the numerical convergence for smaller values of angle $\alpha$, but it is much appreciated for general purpose description. The purpose of this piece of work is to find out the influence of the forces that controls the ion dynamics in the sheath in the presence of inclined magnetic field via the second approach discussed above. Numerical simulation is performed to find the ion dynamics in the plasma sheath by using fluid equations for the ions. These equations are coupled with the Boltzmann relation for the electrons and Poisson's equations. The study will help to acquire some insights on the physics for different dynamic phenomena caused by ions in the presence of magnetic field.

Moving forward, in the following, Sec. II discusses the basic equation and modeling the problem, Sec. III deals with the numerical execution and default parameters, and Sec IV takes care of the results and discussions. The paper is finally concluded in the Sec. V.

## II. Basic equations and modeling

In the model (Figure 1) it has been considered that plasma is in contact with an infinite surface along the x-axis and it has reached the steady state. The surface is assumed to be completely absorbent for any incoming particles. It is assumed that the physical parameters of the sheath vary only along the perpendicular direction (z-direction) to the surface. That means the study has been done in one-dimensional coordinate space and three-dimensional velocity space. This assumption is well known in sheath study[1,7,9,19]. As depicted in Figure 1, the wall is considered on the right-hand side, and the sheath edge is considered on the left-hand side. A steady magnetic field has been applied externally in the x-z plane which makes an angle $\alpha$ with the x-axis. The ions are described by the standard fluid equations. The electrons are assumed to be strongly magnetized, and their guiding centers follows the lines of magnetic field all the way to the wall. Thus, the electrons may be described by Boltzmann distribution. The continuity equation and momentum transport equations for ions are as follows

$$\frac{\partial}{\partial z}(n_i v_z) = S_i \tag{1}$$

$$m_i n_i \left( v_z \frac{d\mathbf{v}}{dz} \right) = -n_i e \frac{d\phi}{dz} \hat{k} + (\mathbf{v} \times \boldsymbol{\omega}) - m_i \mathbf{v} S_i \tag{2}$$

$S_i$ is the ion source term and written as

$$S_i = \begin{cases} \dfrac{n_0}{\tau} \exp\left(\dfrac{e\phi}{kT_e}\right) = n_e Z & \text{(Exponential case)} \\ \dfrac{n_0}{\tau} \cos\left(\dfrac{\pi}{2}\dfrac{v_z}{c_s}\right) H(c_s - v_z) & \text{(Heaviside case)} \end{cases}, \quad Z = 1/\tau \tag{3}$$

The Boltzmann relation for electrons

$$n_e = n_0 \exp\left(\frac{e\phi}{kT_e}\right) \tag{4}$$

The Poisson equation

$$\frac{d^2\phi}{dz^2} = e(n_e - n_i)/\varepsilon_0 \tag{5}$$

The meaning of the symbols is standard. $n_i$ is the ion density, $n_e$ is the electron density, $n_0$ is the equilibrium density, $m_i$ is the ion mass, $\mathbf{v}$ is the average ion velocity having three component $v_x, v_y$ and $v_z$ in x, y, z-direction respectively, $e$ is the elementary charge, $\phi$ is the potential, $T_e$ is electron temperature, $\boldsymbol{\omega}$ is angular frequency having three component $\omega_x, \omega_y$ and $\omega_z$ respectively, $\hat{k}$ is the unit vector in depth direction, $\tau$ is the average time between two consecutive ionizations. $H$ represents the Heaviside function.

From the survey of the literature, it has been observed that there are two types of the source that is popular with this kind of scenario. First, one is called exponential ion source term[14,20–23] and the second one is known as Heaviside source or cosine source. The first kind of source term is taken in a scenario where neutrals are assumed to be ionized due to collision with electrons, and the ionization is proportional to the electron density. The second kind is a bit artificially remodeled. It controls the ionization with the increase in velocity. As soon as it reaches acoustic speed the ionization becomes zero.

Considering three components of velocity momentum transport Eq. (2) can be written into following equations:

$$v_z \frac{dv_x}{dz} = v_y \omega_z - \frac{n_e}{n_i} v_x Z \tag{6}$$

$$v_z \frac{dv_y}{dz} = v_z \omega_x - v_x \omega_z - \frac{n_e}{n_i} v_y Z \tag{7}$$

$$v_z \frac{dv_z}{dz} = -\frac{e}{m_i} \frac{d\phi}{dz} - v_y \omega_x - \frac{n_e}{n_i} v_z Z \tag{8}$$

The system of equations (1) and (4)-(8) has to be normalized with suitable scaling parameters to perform the numerical calculation. The parameters used to normalize in this work are as follows:

$$u = v_x/c_s, v = v_y/c_s, w = v_z/c_s, \eta = e\phi/T_e,$$

$$\xi = z/\lambda_D, \frac{d}{dz} = \left(\frac{1}{\lambda_D}\right)\frac{d}{d\xi}, N_i = n_i/n_0, N_e = n_e/n_0$$

where, $c_s$ is the ion sound speed, $\lambda_D$ is electron Debye length, $u, v, w$ are normalized velocity components, $\eta$ is normalized potential, $\xi$ is normalized distance, $N_i$ and $N_e$ are normalized ion and electron density.

After normalization the system of equations mentioned above can be written as:

$$N_e = \exp(\eta) \tag{9}$$

$$\frac{d^2\eta}{d\xi^2} = \left(\exp(\eta) - N_i\right) \tag{10}$$

$$\frac{\partial}{\partial \xi}(N_i w) = \beta \exp(\eta) \tag{11}$$

$$\frac{du}{d\xi} = \gamma_z \left(\frac{v}{w}\right) - \beta \left(\frac{\exp(\eta)}{N_i}\right)\left(\frac{u}{w}\right) \tag{12}$$

$$\frac{dv}{d\xi} = \gamma_x - \gamma_z \left(\frac{u}{w}\right) - \beta \left(\frac{\exp(\eta)}{N_i}\right)\left(\frac{v}{w}\right) \tag{13}$$

$$\frac{dw}{d\xi} = -\left(\frac{1}{w}\right)\frac{d\eta}{d\xi} - \gamma_x \left(\frac{v}{w}\right) - \beta \left(\frac{\exp(\eta)}{N_i}\right) \tag{14}$$

where, $\beta = \frac{\lambda_D}{c_s} Z = \left(\frac{\varepsilon_0 m_i}{n_0 e^2}\right)^{1/2} Z$, $\gamma_x = \left(\frac{\lambda_D}{c_s}\omega_x\right) = \left(\frac{\varepsilon_0}{n_0 m_i}\right)^{1/2} B_0 \cos\alpha$, $\gamma_y = \left(\frac{\lambda_D}{c_s}\omega_y\right) = 0$, and

$$\gamma_z = \left(\frac{\lambda_D}{c_s}\omega_z\right) = \left(\frac{\varepsilon_0}{n_0 m_i}\right)^{1/2} B_0 \sin\alpha$$

Using Eq. (14) in Eq. (11) we get

$$\frac{dN_i}{d\xi} = \frac{N_i}{w^2}\left(\frac{d\eta}{d\xi}\right) + N_i \gamma_x \left(\frac{v}{w^2}\right) + 2\frac{\beta}{w}\exp(\eta) \tag{15}$$

From equations (10), (12) - (14) and (15) the electrostatic potential, flow velocity field of ions, the electron-ion density distribution can be obtained. Multiplying equations (12) -(14), respectively, by $u, v, w$ then adding and integrating, the following relation is obtained. This relation is also known as energy conservation law.

$$u^2 + v^2 + w^2 = -2\eta + c^2 \tag{16}$$

where $c$ is the integration constant. Using the following conditions $\xi \to 0$, $u = 1$, $v = 0$, $w = 0$, $\eta \to 0$, $c$ is found to be unity.

## III. Numerical execution

Equations(10), (15) and (12) – (14) are solved numerically by Runge-Kutta fourth order method. As the problem is considered as an initial value problem, it is required to provide some initial physical value for each differential equation. The values used to solve these equations are as follows.

$$\eta|_{\xi=0} = 0, \ E|_{\xi=0} = 0.01, \ N_i|_{\xi=0} = 1, \ u|_{\xi=0} = v|_{\xi=0} = 0, \ w|_{\xi=0} = 1$$

The assumption of the initial values was inspired by the following works[9,11]. The ionization frequency is considered as $Z = 10^5 \ s^{-1}$ and equilibrium density as $n_0 = 10^{16} \ m^{-3}$.

## IV. Results and discussions

### A. Variation of the angle $\alpha$

In this paper, the electric field is along the z-direction, and the ions are intended to move towards the wall under the combined action of the electric and the magnetic field. Let us first see how the angle variation affects the whole dynamics of ions. Figure 2 and 3 shows the evolution of the electric potential and the electric field along the z-direction. The idea of sheath thickness can be inferred from these figures. The sheath seems to decrease in length with the decrease in the angle ($\alpha$). The result can be explained from the figure 4, which shows the space charge evolution inside the sheath region. $\alpha = 89^0$ implies that the magnetic field is almost parallel to the z-axis. Thus, the ions have an increasing tendency to move towards the wall. Therefore, at this angle, the sheath formation is similar in nature to that of the electrostatic case. But, as soon as $\alpha$ starts decreasing, spaces charges develop highly within a narrow range of space. This results in the attainment of screening over a smaller length as compared to a higher $\alpha$.

Figure 5 and 6 shows the evolution of the ion density and velocity over the range of angles. The density plot too shows that the sheath formation is alike in the electrostatic case at higher angles. Ion peaks develop gradually with lowering the value of $\alpha$. Moreover, for higher angles, the B and E are almost parallel and hence, the ions move with increasing velocity along z. The other two components are found to be zero for $\alpha = 89^0$. However, with the increase in the angle, magnetic field starts affecting the situation, and the other two component of velocity, i.e., $u$ and $v$ share the net velocity at a location with $w$. For the extreme low case, $\alpha = 45^0$ the $u$ component competes with $w$ significantly.

Figure 7 shows a three-dimensional plot of the $u, v$ and $w$. A helix is seen to evolve even in the 3D velocity space. The number of turns of the helix decreases with decreasing $\alpha$. For higher angles, the helix opens up from the above and hardly completes a single turn thereafter. This clearly shows that very near to the wall, for higher angles the $w$ component dominates and the other two losses significance.

One important parameter in the magnetic geometry has been the calculation of the pitch angle, which is the angle between the particle velocity vector and the magnetic field. The angle is given by,

$$\theta = \tan^{-1}\left(\frac{v_{perp}}{v_{para}}\right)$$

Here, $v_{perp} = \sqrt{(w\cos(\alpha) - u\sin(\alpha))^2 + v^2}$ and $v_{para} = u\cos(\alpha) + w\sin(\alpha)$.

The perpendicular component $v_{perp}$ is responsible for the rotation of particles around the magnetic field and the parallel component ($v_{para}$) is responsible for giving it a leap in the direction of the magnetic field. Figure 8 shows the plot of the pitch angle with distance along *z*. The figure shows very regular periodic patterns for higher angles, but with lowering the angle, the regular periodic behavior is destroyed. It is very interesting to note that the maximum amplitude happens to the angle (90-$\alpha$) i.e. the angle between the magnetic field and the *z*-axis.

Figure 9 portrays the pitch length for different angle variations. In general, the pitch length is calculated from the relation

$$P = v_{para} T$$

Here, T is the time period of revolution of the particle around the magnetic field. In the normalized order, this is given by,

$$P' = \left(\frac{C_s T}{\lambda_D}\right) v_{para}$$

Thus, pitch length is proportional directly to the parallel component of the velocity. While the particle rotates around the magnetic field, the pitch length tells how much the particle has leaped forward. In the present paper, the exact particle trajectory is unknown and rather the velocity field construct is available with components. Therefore, the pitch length here says that $P'$ would be the pitch length if the particle were allowed to rotate in a constant magnetic field with velocity $v_{para}$. But, since because of the action of the electric field the parallel velocity changes hence a continuous evolution of pitch length is obtained. The evolution, however, shows the length to increase along *z*. For the higher angle, the pitch length remains by far steady initially and then near to the wall starts increasing. For the lower angle, this is seen to increase from the very beginning.

**B. Variation of the magnetic field strength B**

The similar set of profiles as discussed above are obtained with the variation of the strength of the magnetic field, keeping the angle fixed at $45^0$. The field is varied from 1T to 4T. Figure 10 and 11 shows the variation of the electric potential and the electric field for the field strength variation. Figure 12 shows the space charge evolution throughout the domain. It is interesting that even with increasing the strength of the magnetic field the sheath thickness is decreased. This is once again for the high space charge deposition over a narrow region of space. The assimilation of high space charges is also seen in the density plot in Figure 13. Figure 14 shows the velocity evolution over the domain. For low magnetic field (1T), *u* and *v* component goes parallel to each other along *z*, and the *w* component is higher than those. However, at the magnetic field of 4T, *u* component is seen to surpass the *w* component at the wall.

Figure 15 and 16 shows the plot in 3D for *u*, *v* and *w* component of velocity and the pitch angle variation along *z*. For low magnetic field, as it is seen in figure 15, the $(u,v,w)$ plot hardly completes a revolution in the plane. However, with the increase of the magnetic field strength, the tendency increases to complete a rotation. This implies that with the increase in field strength, *u* and *v* component gets more share of the velocity over *w*. Out of these three, the *v* component is responsible for rotating the particles in its plane of rotation with the increased radius. This happens because of the angular frequency ($\Omega_c = \frac{qB}{m}$) of rotation of the particle. The $\Omega_c$ is directly proportional to the strength of the magnetic field (B). This, in turn, increases the frequency of rotation. Thus more of the kinetic energy would be used up in rotating than in going parallel to B. This enhances the $v$ component of the velocity. Figure 17 shows the variation of the pitch length along $z$ for different magnetic field strengths.

### C. Discussion on the Lorentz force

Figure 18 shows the plot of (*x, y, z*) component of magnetic force and the electrostatic force along with the net force along the *z*-direction. The *x* component of the magnetic force shows a local maximum initially and rises breaking all bounds thereafter. The *z*-component of the force field is negative initially and rises to a positive value for some region of space and again falls off to a negative value. Therefore, the electrostatic force is solely responsible for creating a velocity field along *z*.

This should be noted that despite the net force along *z* being negative, the *w* component is positive always. This is because the ions enter the sheath with a definite velocity called the Bohm velocity along *z*. This gives rise to the required kinetic energy to prosper along *z* despite adversity. The initial zone of the sheath is very critical for this reason. For the formation of the sheath, the electric field must support the velocity along *z* before it falls off to zero. Comparing the velocity plots and the net force along *z* we see that this criticality is maintained by the electric field and sheath forms. Wherever the net force along *z* is zero, exactly there the *w* component reaches maximum deceleration. From the very next point the force is positive (because of the electric field), and the *w* component rises again. This criticality is present in all cases irrespective of the angle variation or the field strength variation.

### D. Discussion on the energy of the ions

Figure 19 shows the total energy and the component energies to vary along *z* for a wide range of magnetic field at $\alpha = 45^0$. For higher field (4T) the major part of the total energy is contributed by the *x* and *z* component of velocity. The *y* component has some contribution to it towards the wall only. The interesting fact is that the *x* component is even higher than the *z* component at the wall. For all other values of the magnetic field, the *z* component dominates over the other two.

Figure 20 shows the plot of energy with variation in the angle $\alpha$. The magnetic field is kept fixed at 4T. The figure is quite unlike the previous one. Here with the increase in angle, the *z* component of energy fully dominates over the other two, and towards the higher angle, the whole contribution to the energy come from this only. The contribution of the *y* component is once again insignificant even at lower angles.

### E. Discussion on the source terms of the ions

Source terms can create a big difference in the results according to the scenario. Figure 21 shows that for smaller incident angle Heaviside source pulls down the ion density as the source restricts ionization when ion velocity reaches acoustic speed. Even the flow velocity, as well as pitch, becomes out of phase at an early stage (see figure 22). In another scenario at a higher angle $\alpha = 89^0$ both the ion source terms behaves almost in a similar fashion (see figure 23). In figure 24 in the same scenario, the flow velocity and pitch also shows the same fact. Although they become out of phase at a later stage near the wall as the flow velocity reaches acoustic speed.

## V. CONCLUSIONS

This paper provides the insight on the effect of force fields on ion dynamics in a sheath in the presence of an inclined magnetic field. The study of particle dynamics is always interesting as it unfolds much information on any plasma system. The results clearly support the earlier fact that the sheath width gets smaller and smaller as the inclination angle starts decreasing. From the electric field profile, it has been observed that it is getting dispersed with the variation of $\alpha$. The dispersion of the electric field with decreasing $\alpha$ is quite important when the consequences on the recycling of sputtered particles in a tokamak edge are considered. There is a scope to study the effect of the fact mentioned above. The development of peak in ion density distribution with lowering the value of $\alpha$ certainly indicates the deceleration of ions due to magnetic force towards the wall. The breaking of the periodicity of pitch angle due to smaller inclination angle gives a new perception of the ion flow inside the sheath. Although the magnetic field is considered to be uniform throughout the domain, the pitch length depicts a different picture from a usual one. The analysis of Lorentz force helps a lot to understand and explain the dynamic behavior of the ions in the system. The effect of the magnetic field strength also shows some interesting behavior while studying. While kinetic analysis gives a complete picture of particle dynamics, the present fluid analysis provides a clear overall understanding of such systems. In conclusion, this paper opens up a new doorway to understanding the ion dynamics inside the sheath in the presence of magnetic field through fluid approach.

**ACKNOWLEDGMENTS:**

The authors wish to thank Prof. P. K. Kaw of Institute for Plasma Research, Gandhinagar, India for his idea to conduct this study.

**REFERENCES:**

[1] K.U. Riemann, Phys. Plasmas **1**, 552 (1994).

[2] S. Devaux and G. Manfredi, Plasma Phys. Control. Fusion **50**, 25009 (2008).

[3] Roshan Chalise and Raju Khanal, J. Mater. Sci. Eng. A **5**, 41 (2015).

[4] S. Kuhn, D.D. Tskhakaya, and D. Tskhakaya, Comput. Phys. Commun. **177**, 80 (2007).

[5] J. Kovačič, T. Gyergyek, and M. Čerček, Eur. Phys. J. D **54**, 383 (2009).

[6] D.L. Holland, The Magnetized Plasma Sheath, University of California, Los Angeles, 1990.

[7] R. Chodura, Phys. Fluids **25**, 1628 (1982).

[8] D. Tskhakaya and S. Kuhn, J. Nucl. Mater. **313–316**, 1119 (2003).

[9] S. Farhad Masoudi, S.S. Esmaeili, and S. Jazavandi, Vacuum **84**, 382 (2009).

[10] S.F. Masoudi and S.M. Salehkoutahi, Eur. Phys. J. D **57**, 71 (2010).

[11] X. Zou, J.-Y. Liu, Y. Gong, Z.-X. Wang, Y. Liu, and X.-G. Wang, Vacuum **73**, 681 (2004).

[12] J.P. Gunn, Phys. Plasmas **4**, 4435 (1997).

[13] J. Ou and J. Yang, Phys. Plasmas **19**, 113504 (2012).

[14] J. Liu, F. Wang, and J. Sun, Phys. Plasmas **18**, (2011).

[15] T. Gyergyek and J. Kovačič, Phys. Plasmas **22**, 43502 (2015).

[16] D.L. Holland, B.D. Fried, and G.J. Morales, Phys. Fluids B Plasma Phys. **5**, 1723 (1993).

[17] A. Bergmann, Phys. Plasmas **9**, 3413 (2002).

[18] R. Cohen and D. Ryutov, Phys. Plasmas **5**, 808 (1998).

[19] P.C. Stangeby, Nucl. Fusion **52**, 83012 (2012).

[20] K.-U. Riemann, Contrib. Plasm. Phys. **34**, 127 (1994).

[21] R.N. Franklin, J. Plasma Phys. **78**, 21 (2011).

[22] T.M.G. Zimmermann, M. Coppins, and J.E. Allen, Phys. Plasmas **15**, (2008).

[23] J. Ou, N. Xiang, C. Gan, and J. Yang, Phys. Plasmas **20**, (2013).

**CAPTIONS:**

**FIG. 1.** (Color online) Schematic of the magnetic sheath problem (top). Vector diagram of the problem (bottom). **B** is on the $x-z$ plane.

**FIG. 2.** (Color online) The electric potential ($\eta$) profiles at a magnetic field strength ($B$) 4T.

**FIG. 3.** (Color online) The electric field ($E$) profiles at a magnetic field strength ($B$) 4T.

**FIG. 4.** (Color online) The space charge ($\sigma$) profiles at a magnetic field strength ($B$) 4T.

**FIG. 5.** (Color online) The density distributions at a magnetic field strength ($B$) 4T.

**FIG. 6.** (Color online) The ion flow velocity components at a magnetic field strength ($B$) 4T.

**FIG. 7.** (Color online) The ion flow velocities at a magnetic field strength ($B$) 4T.

**FIG. 8.** (Color online) The ion pitch angle profiles at a magnetic field strength ($B$) 4T.

**FIG. 9.** (Color online) The ion pitch length profiles at a magnetic field strength ($B$) 4T.

**FIG. 10.** (Color online) The electric potential ($\eta$) profiles at a magnetic inclination angle ($\alpha$) $45°$.

**FIG. 11.** (Color online) The electric field ($E$) profiles at a magnetic inclination angle ($\alpha$) $45°$.

**FIG. 12.** (Color online) The space charge ($\sigma$) profiles at a magnetic inclination angle ($\alpha$) $45°$.

**FIG. 13.** (Color online) The density distributions at a magnetic inclination angle ($\alpha$) $45°$.

**FIG. 14.** (Color online) The ion flow velocity components at a magnetic inclination angle ($\alpha$) $45°$.

**FIG. 15.** (Color online) The ion flow velocities at a magnetic inclination angle ($\alpha$) $45°$.

**FIG. 16.** (Color online) The ion pitch angle profiles at a magnetic inclination angle ($\alpha$) $45°$.

**FIG. 17.** (Color online) The ion pitch length profiles at a magnetic inclination angle ($\alpha$) $45°$.

**FIG. 18.** (Color online) The Force field profiles at a magnetic field strength ($B$) 4T and magnetic inclination angle ($\alpha$) $45°$.

**FIG. 19.** (Color online) The energy profiles at a magnetic inclination angle ($\alpha$) $45°$.

**FIG. 20.** (Color online) The energy profiles at a magnetic field strength ($B$) 4T.

**FIG. 21.** (Color online) The density distributions at a magnetic field strength ($B$) 4T and at a magnetic inclination angle ($\alpha$) $45°$.

**FIG. 22.** (Color online) The ion flow velocities at a magnetic field strength ($B$) 4T and at a magnetic inclination angle ($\alpha$) $45°$.

**FIG. 23.** (Color online) The density distributions at a magnetic field strength ($B$) 4T and at a magnetic inclination angle ($\alpha$) $89°$.

**FIG. 24.** (Color online) The ion flow velocities at a magnetic field strength ($B$) 4T and at a magnetic inclination angle ($\alpha$) $89°$.

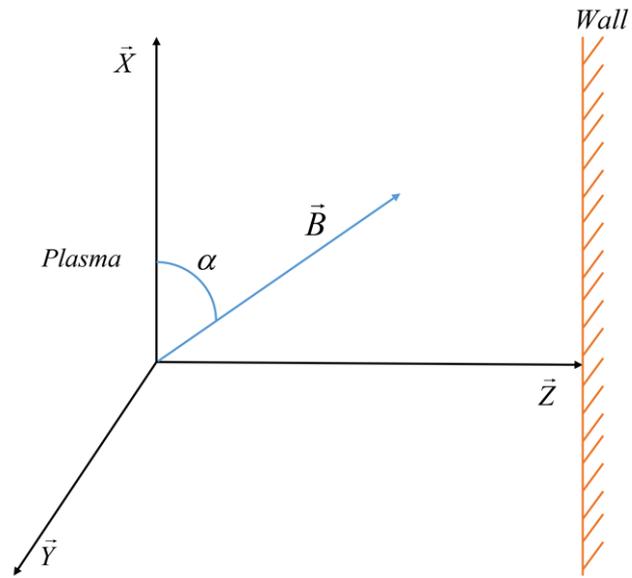

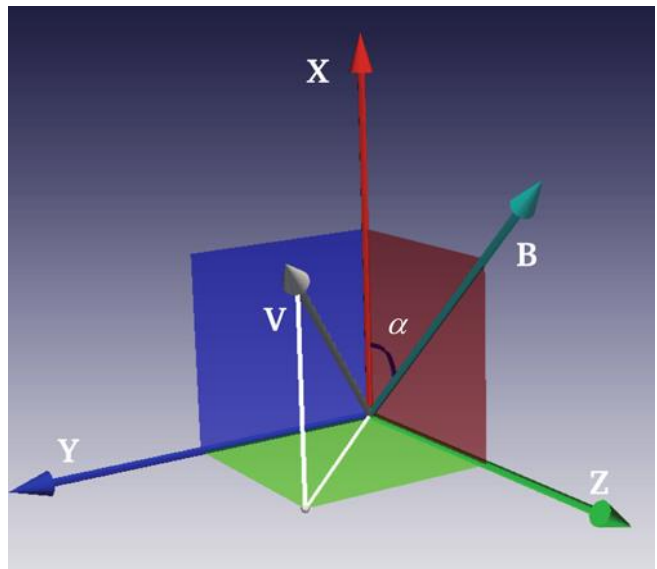

**FIG. 1.** (Color online) Schematic of the magnetic sheath problem (top). Vector diagram of the problem (bottom). $B$ is in the $x-z$ plane.

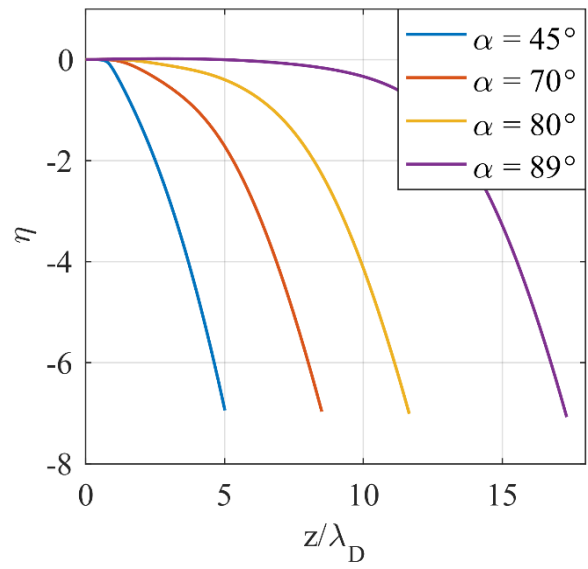

**FIG. 2.** (Color online) The electric potential ($\eta$) profiles at magnetic field strength ($B$) 4T.

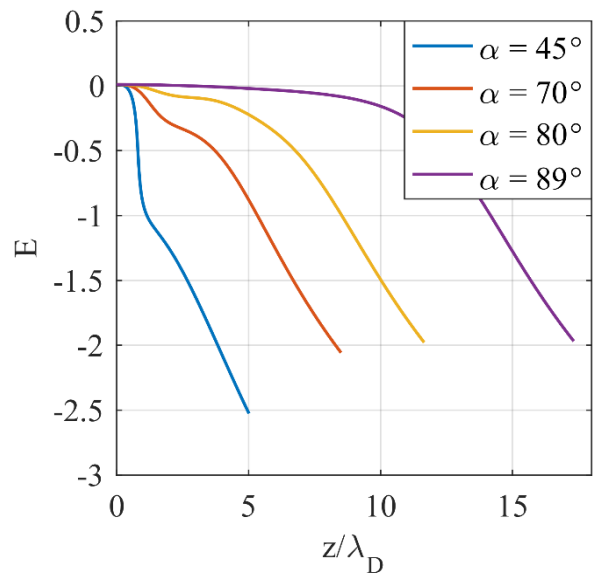

**FIG. 3.** (Color online) The electric field ( $E$ ) profiles at magnetic field strength ( $B$ ) 4T.

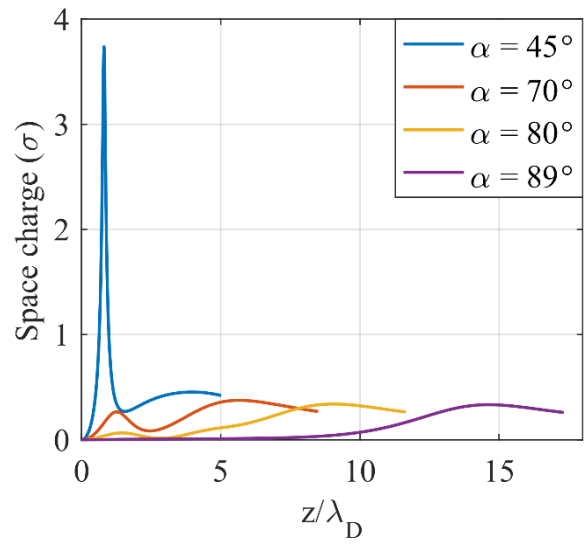

**FIG. 4.** (Color online) The space charge ($\sigma$) profiles at magnetic field strength ($B$) 4T.

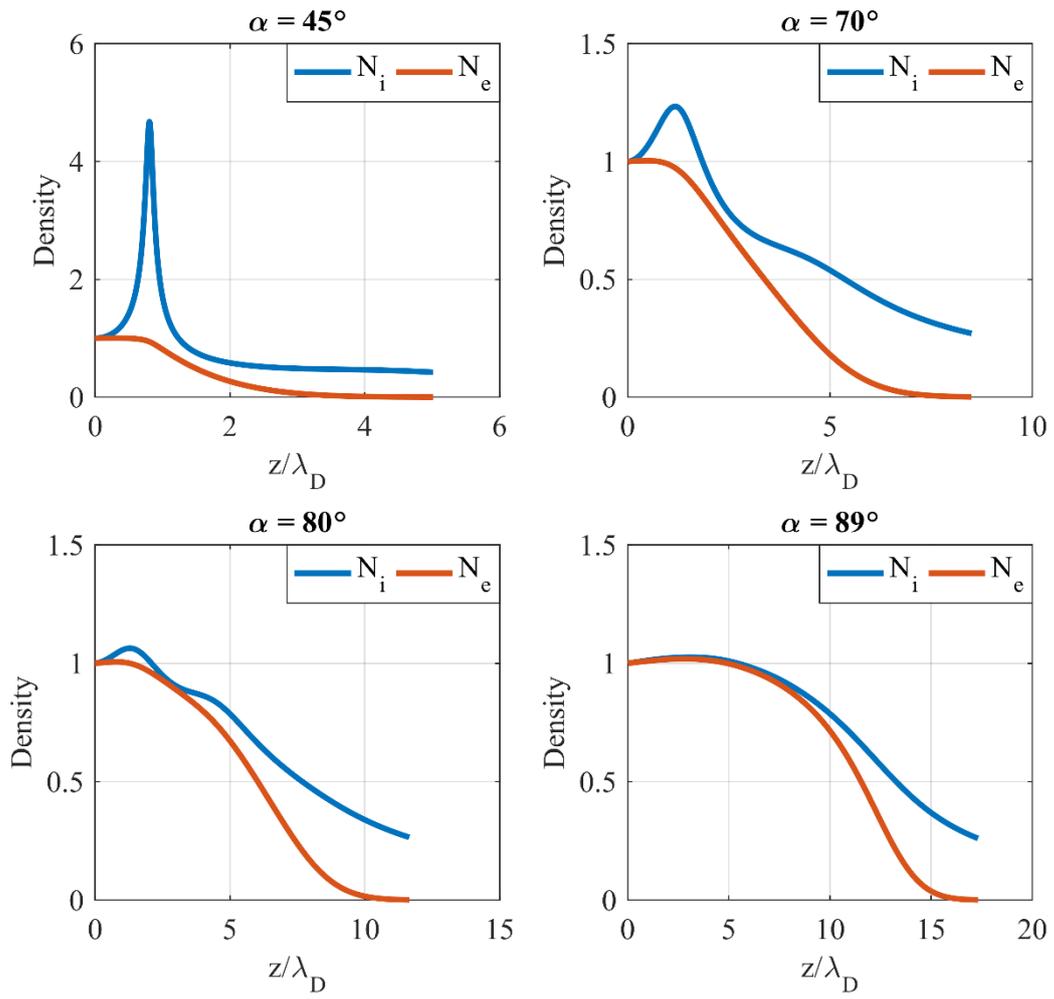

**FIG. 5.** (Color online) The density distributions at magnetic field strength ($B$) 4T.

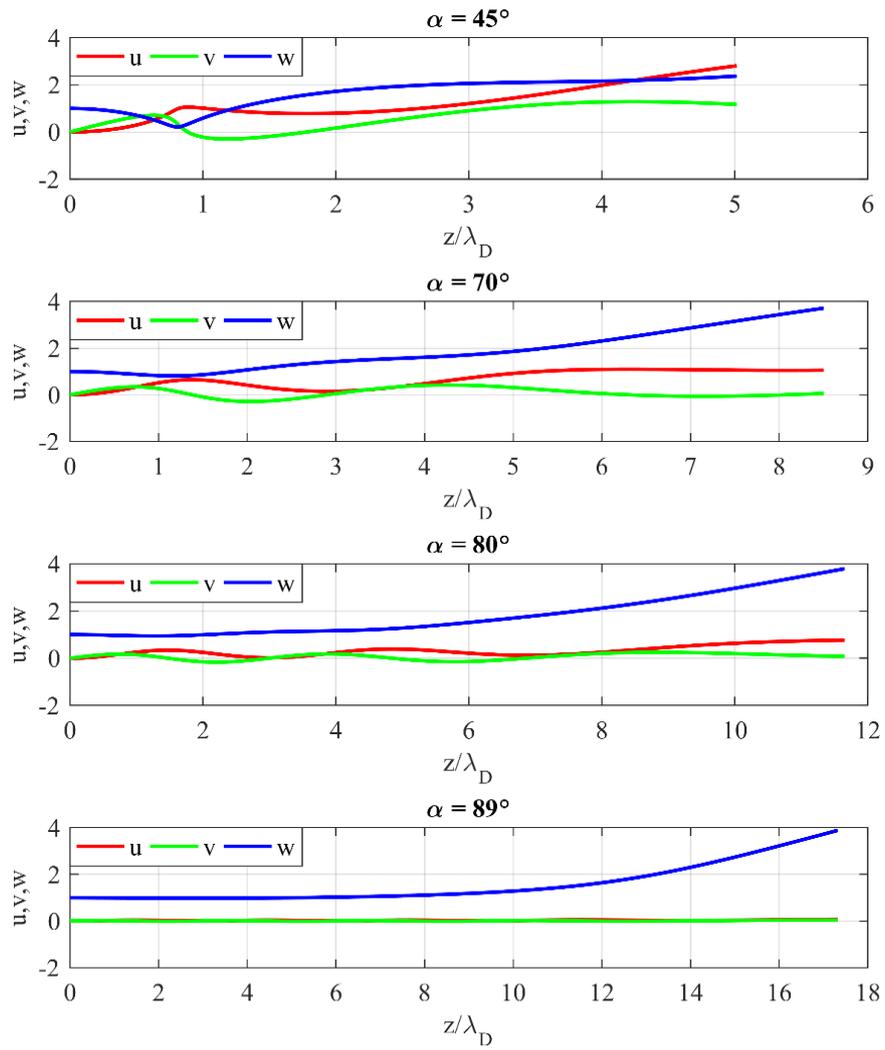

**FIG. 6.** (Color online) The ion flow velocity components at magnetic field strength ($B$) 4T.

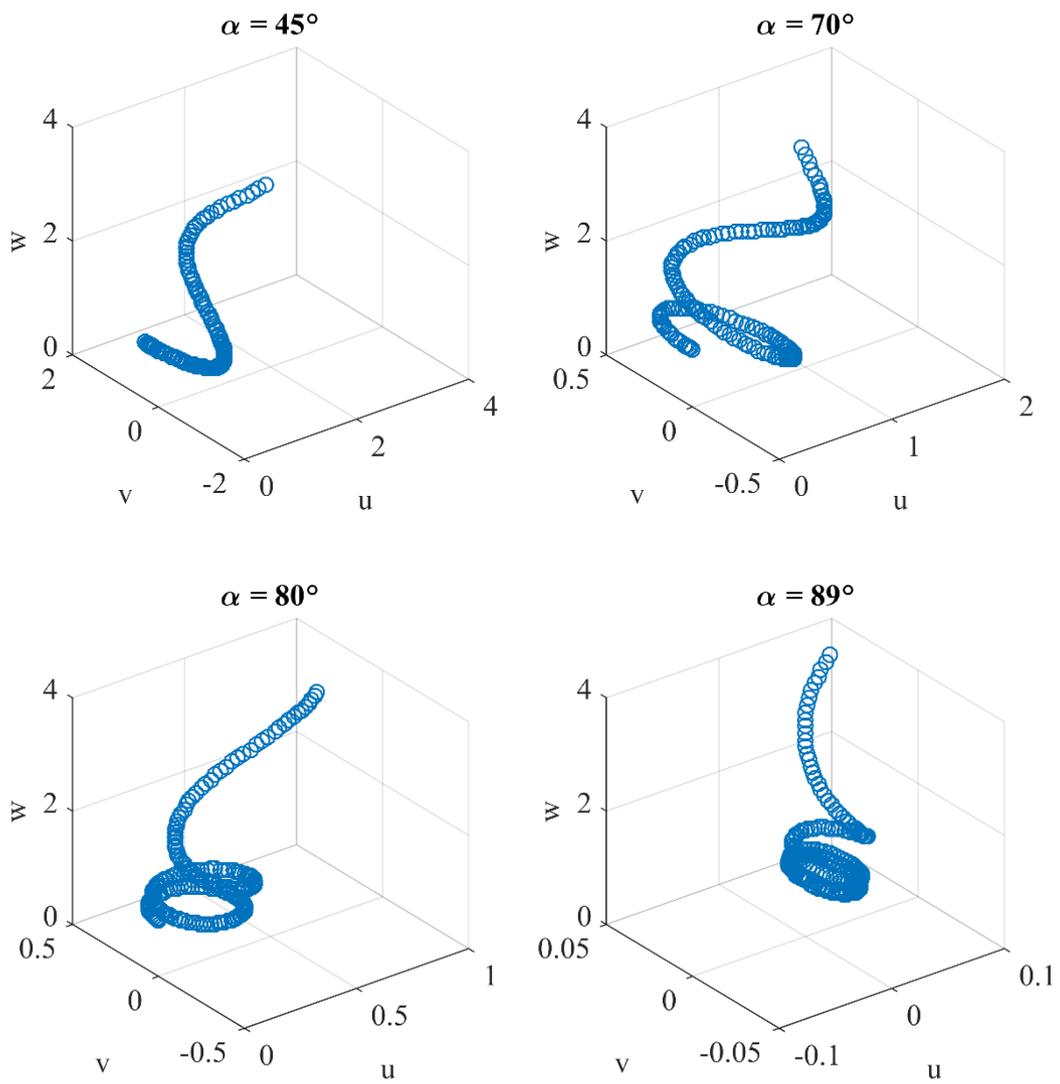

**FIG. 7.** (Color online) The ion flow velocities at magnetic field strength ($B$) 4T.

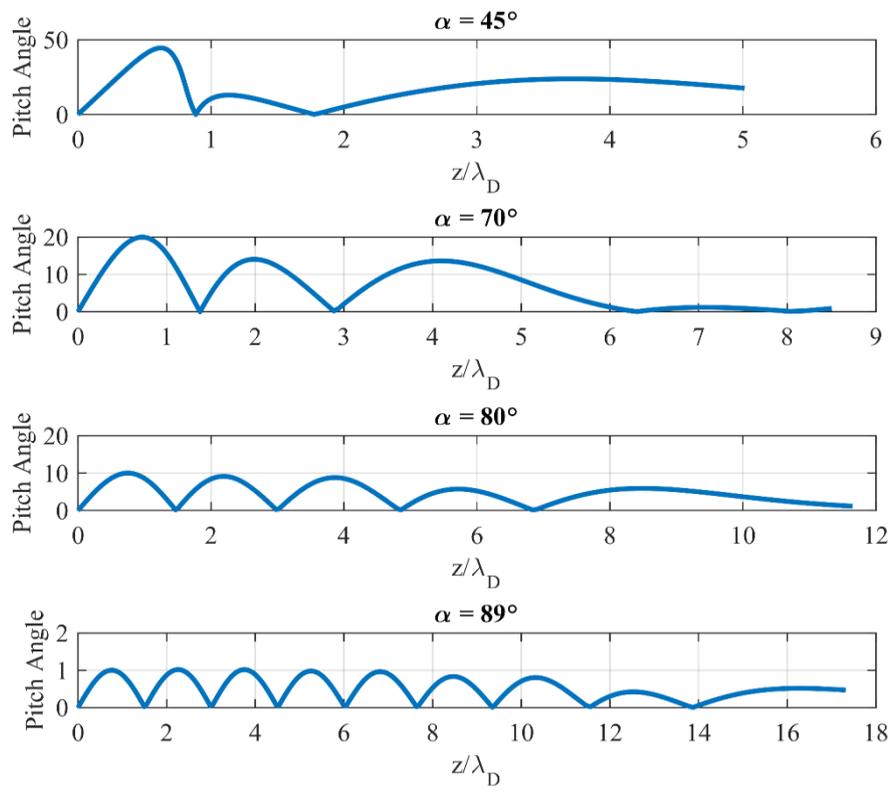

**FIG. 8.** (Color online) The ion pitch angle profiles at magnetic field strength ( $B$ ) 4T.

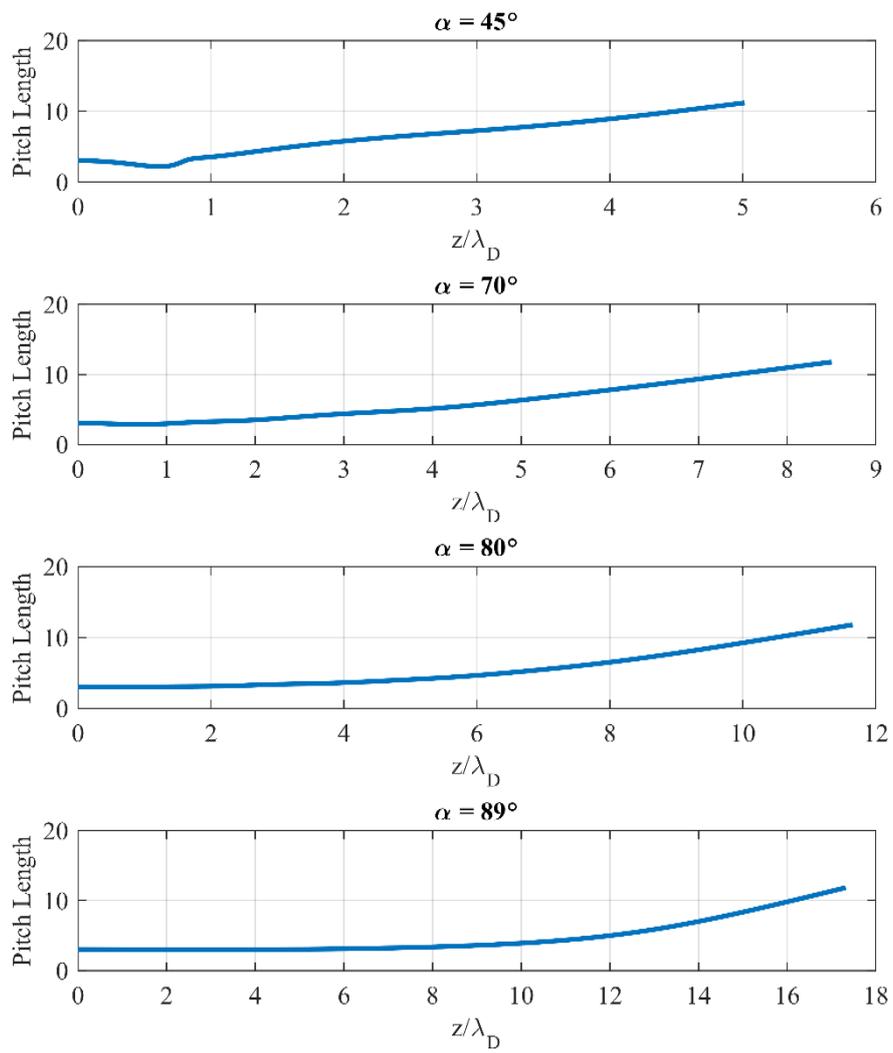

**FIG. 9.** (Color online) The ion pitch length profiles at magnetic field strength ($B$) 4T.

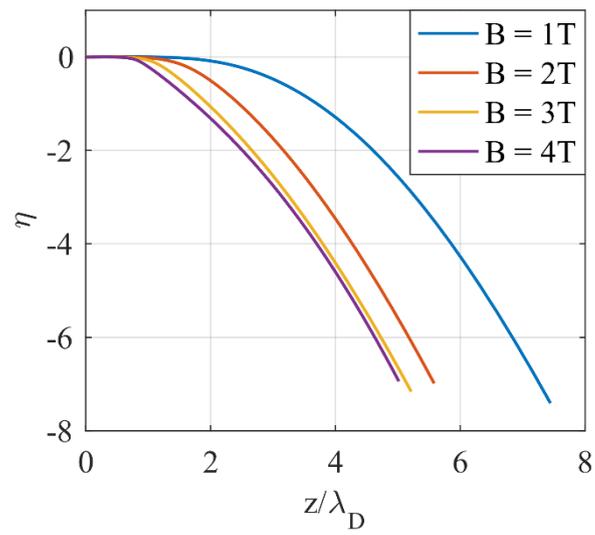

**FIG. 10.** (Color online) The electric potential ($\eta$) profiles at magnetic inclination angle ($\alpha$) 45°.

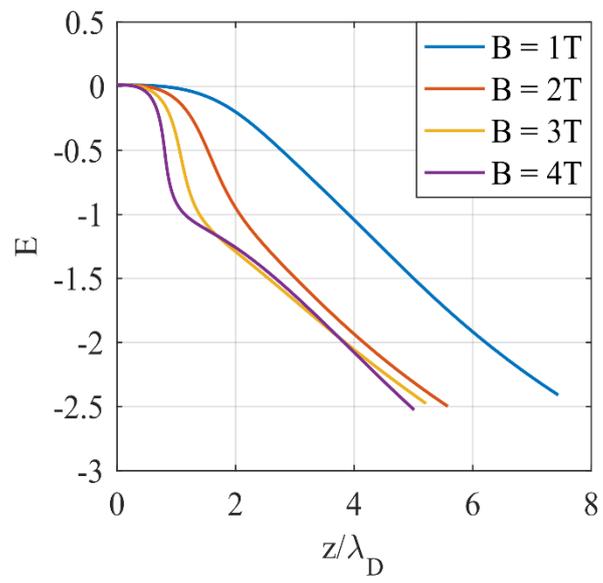

**FIG. 11.** (Color online) The electric field ($E$) profiles at magnetic inclination angle ($\alpha$) $45°$.

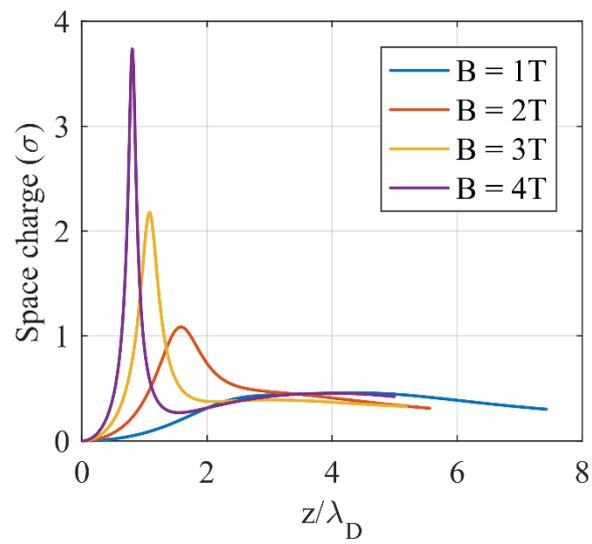

**FIG. 12.** (Color online) The space charge ($\sigma$) profiles at magnetic inclination angle ($\alpha$) 45°.

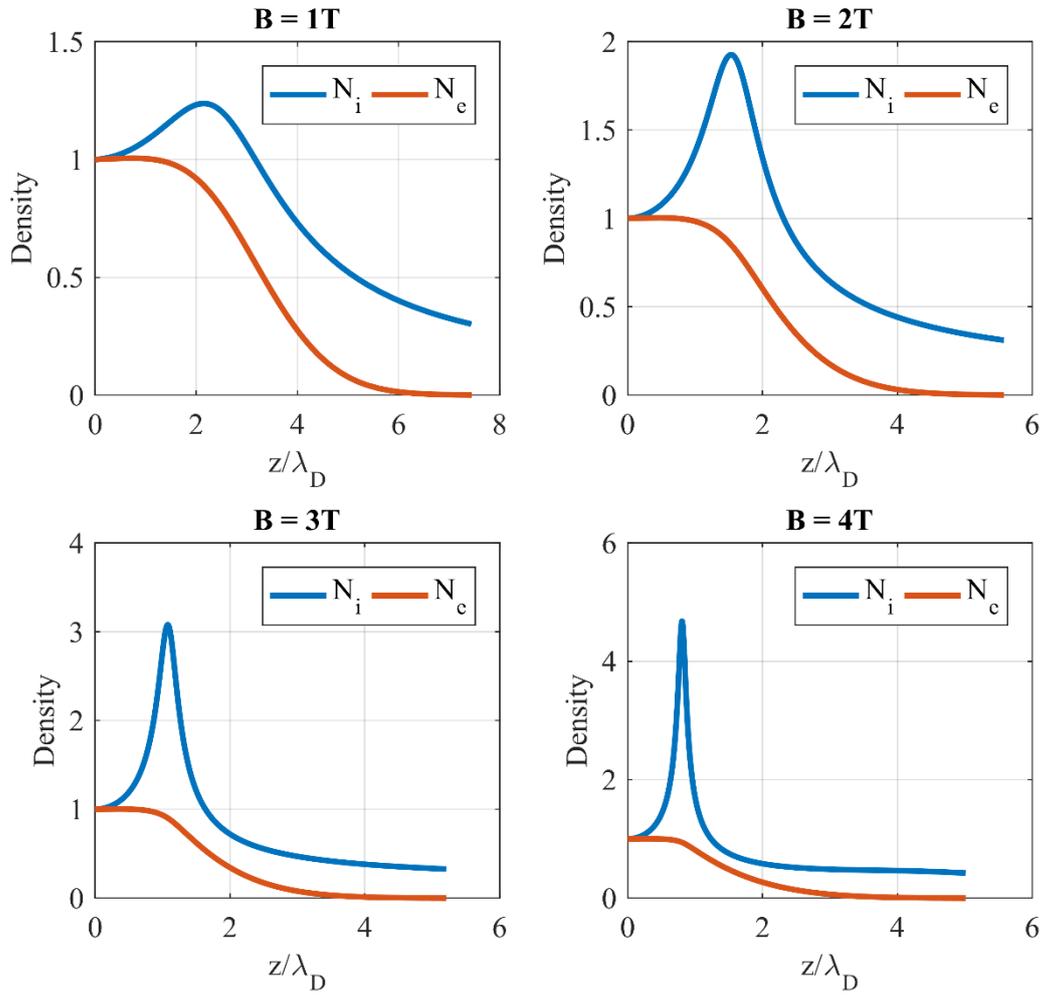

**FIG. 13.** (Color online) The density distributions at magnetic inclination angle ($\alpha$) 45°.

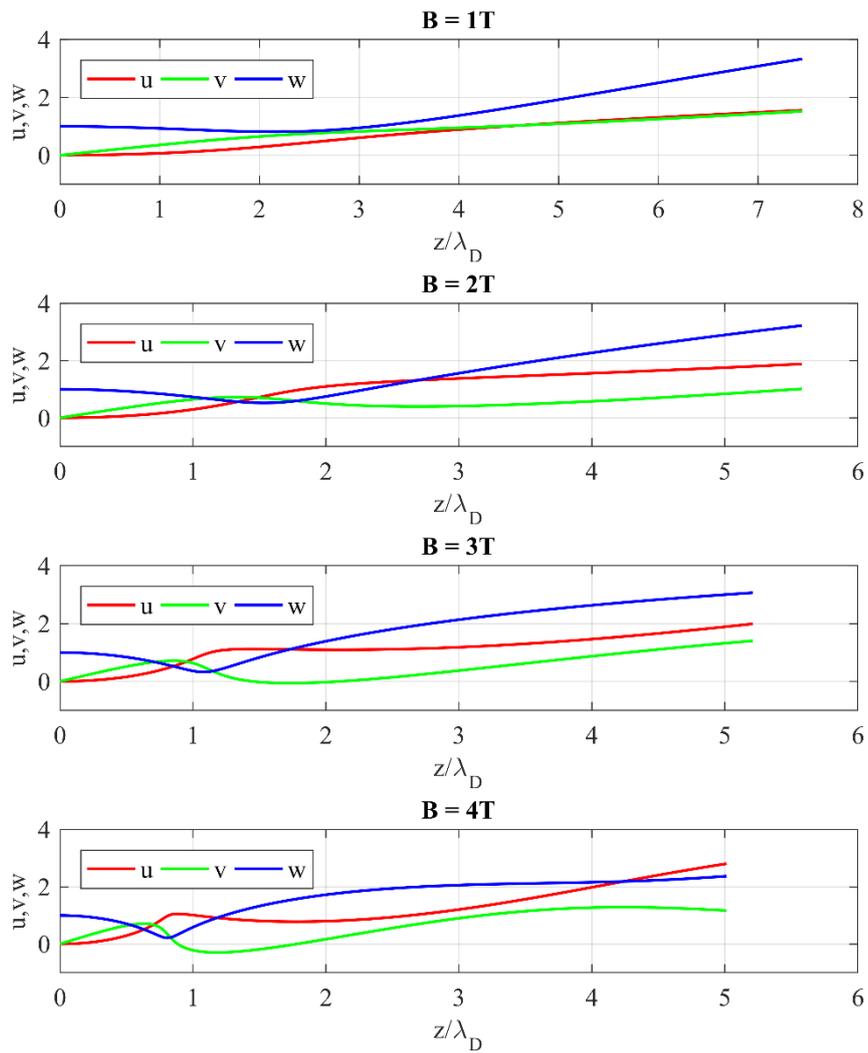

**FIG. 14.** (Color online) The ion flow velocity components at magnetic inclination angle ($\alpha$) $45°$.

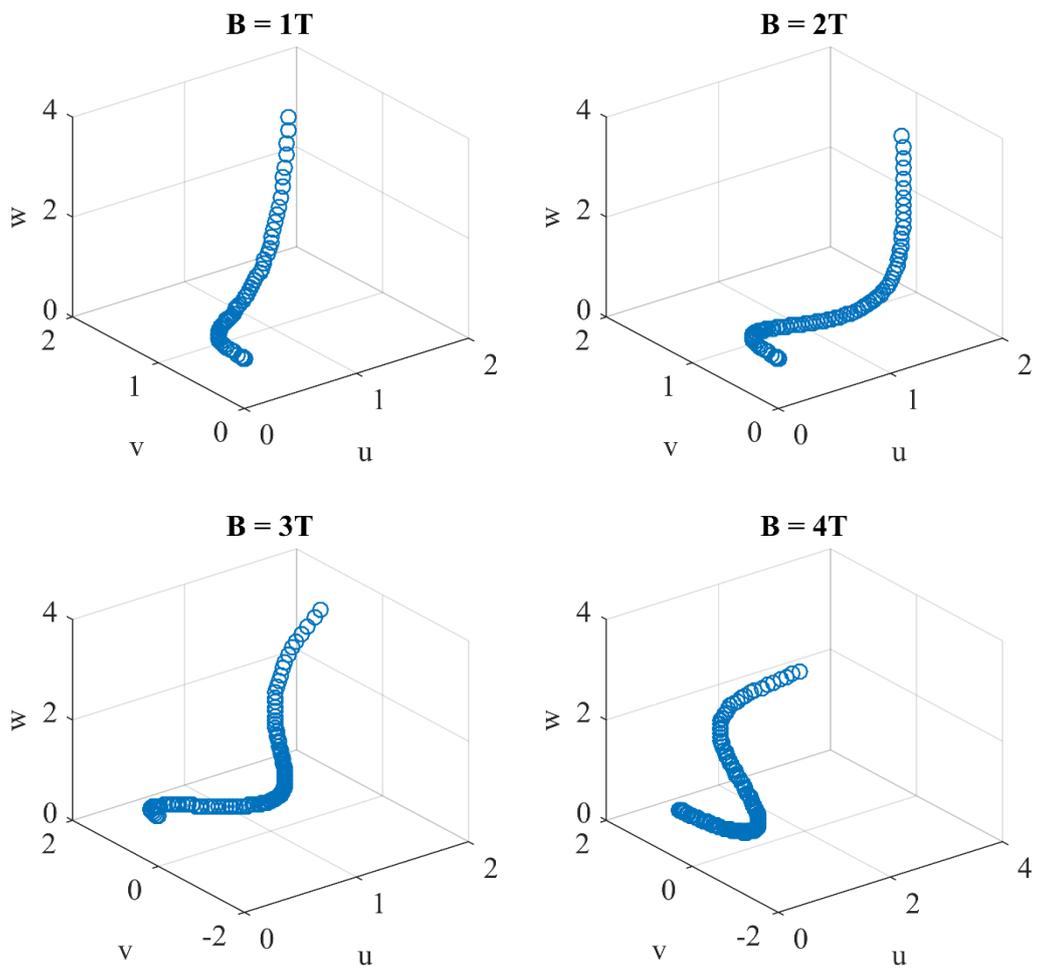

**FIG. 15.** (Color online) The ion flow velocities at magnetic inclination angle ($\alpha$) 45°.

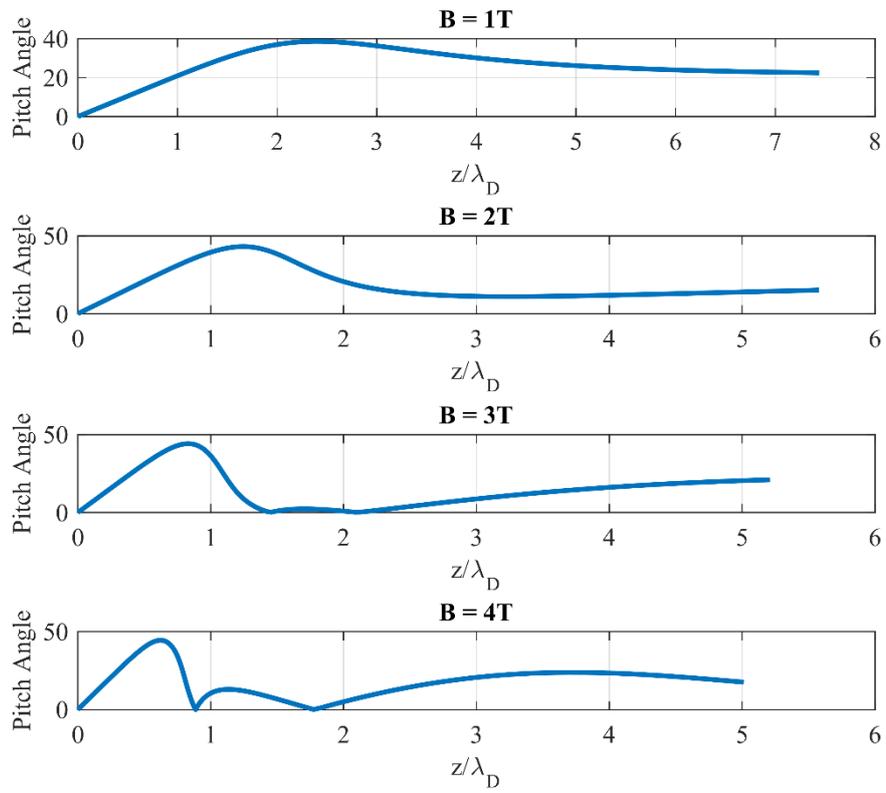

**FIG. 16.** (Color online) The ion pitch angle profiles at magnetic inclination angle ($\alpha$) 45°.

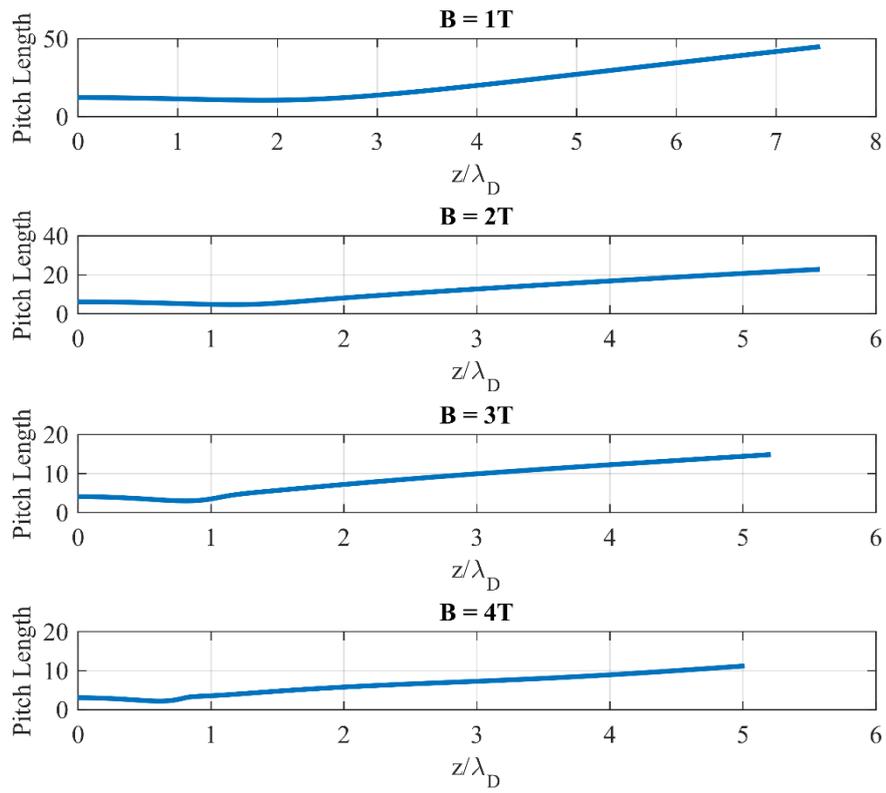

**FIG. 17.** (Color online) The ion pitch length profiles at magnetic inclination angle ($\alpha$) 45°.

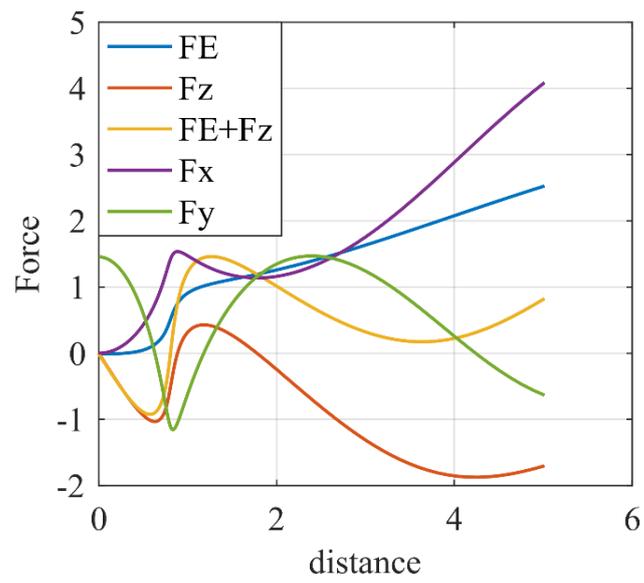

**FIG. 18.** (Color online) The Force field profiles at magnetic field strength ($B$) 4T and magnetic inclination angle ($\alpha$) 45°.

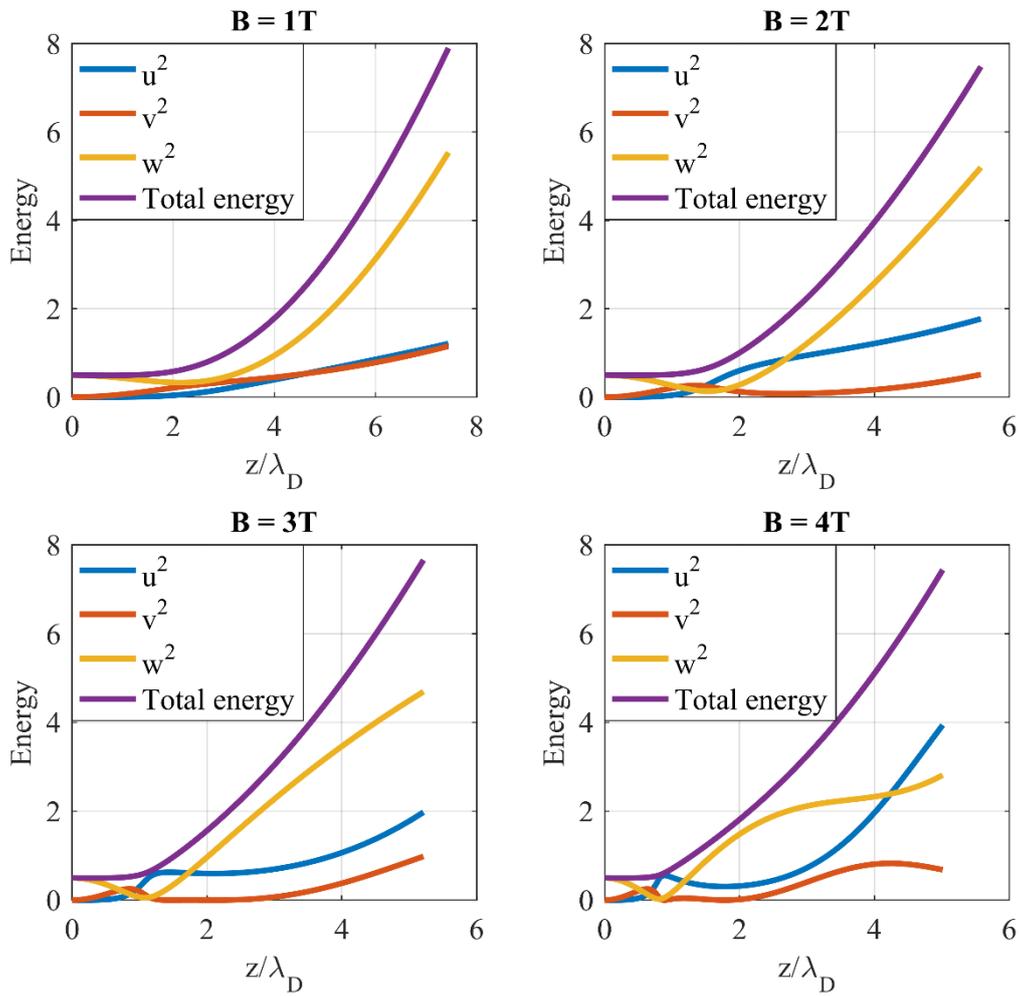

**FIG. 19.** (Color online) The energy profiles at magnetic inclination angle ($\alpha$) 45°.

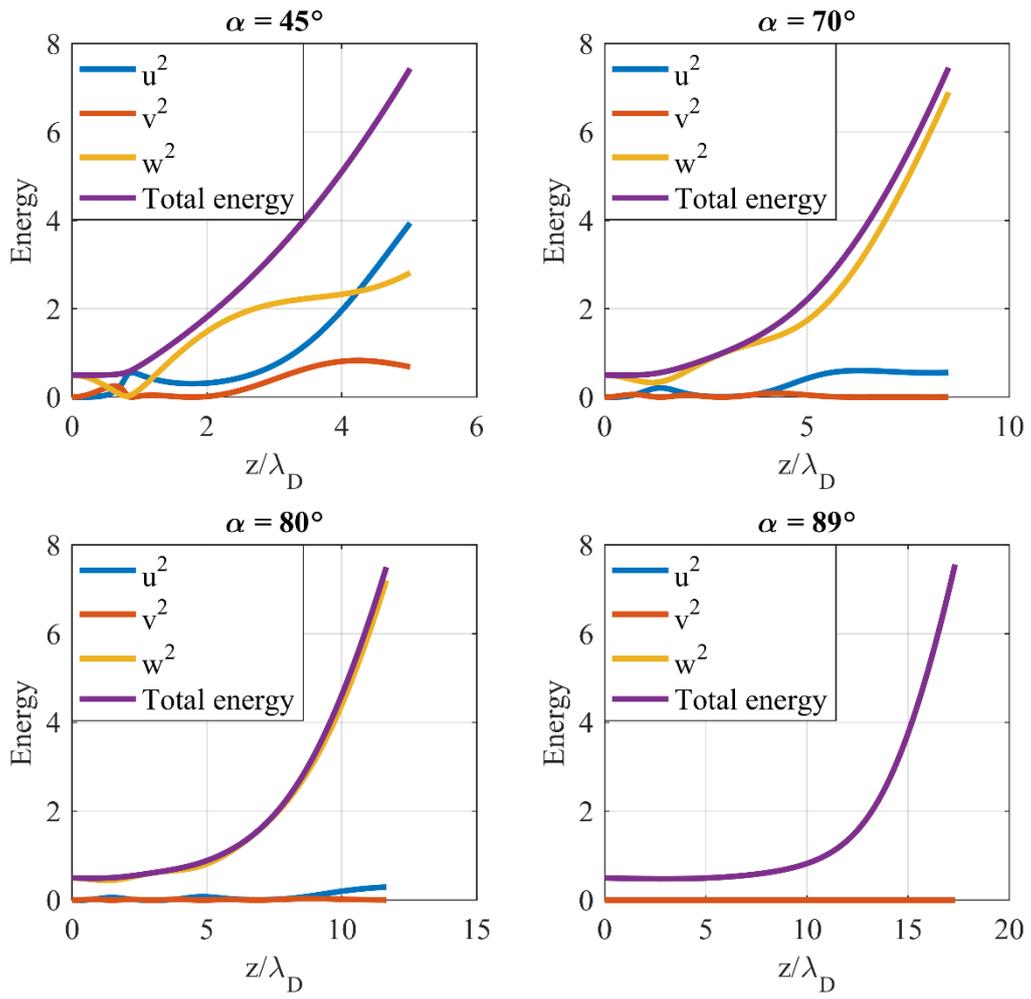

**FIG. 20.** (Color online) The energy profiles at magnetic field strength ( $B$ ) 4T.

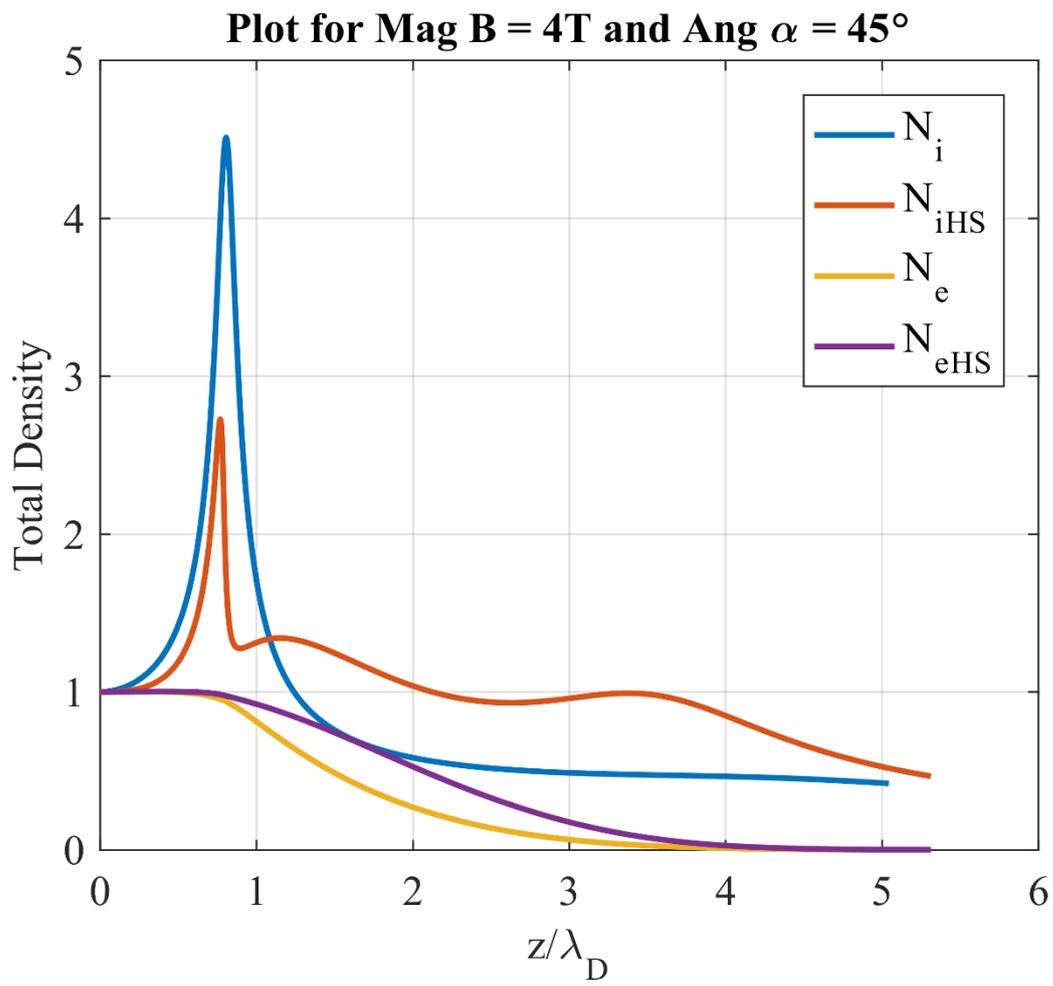

**FIG. 21.** (Color online) The density distributions at a magnetic field strength ($B$) 4T and at a magnetic inclination angle ($\alpha$) 45°.

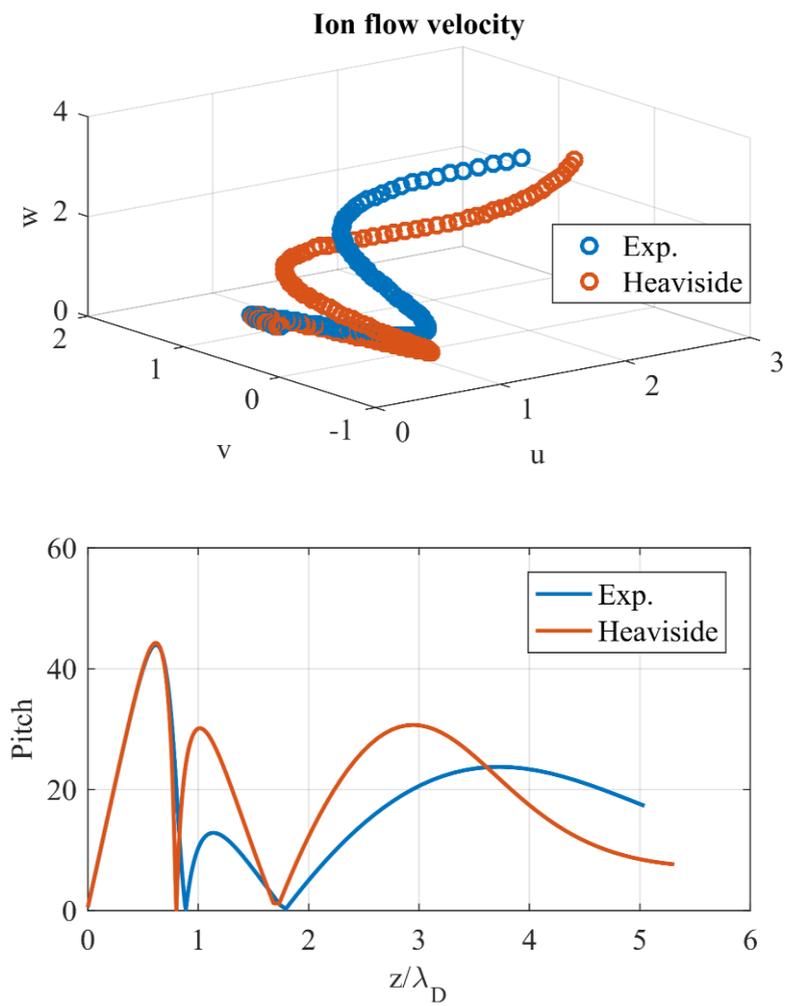

**FIG. 22.** (Color online) The ion flow velocities at a magnetic field strength ($B$) 4T and at a magnetic inclination angle ($\alpha$) 45°.

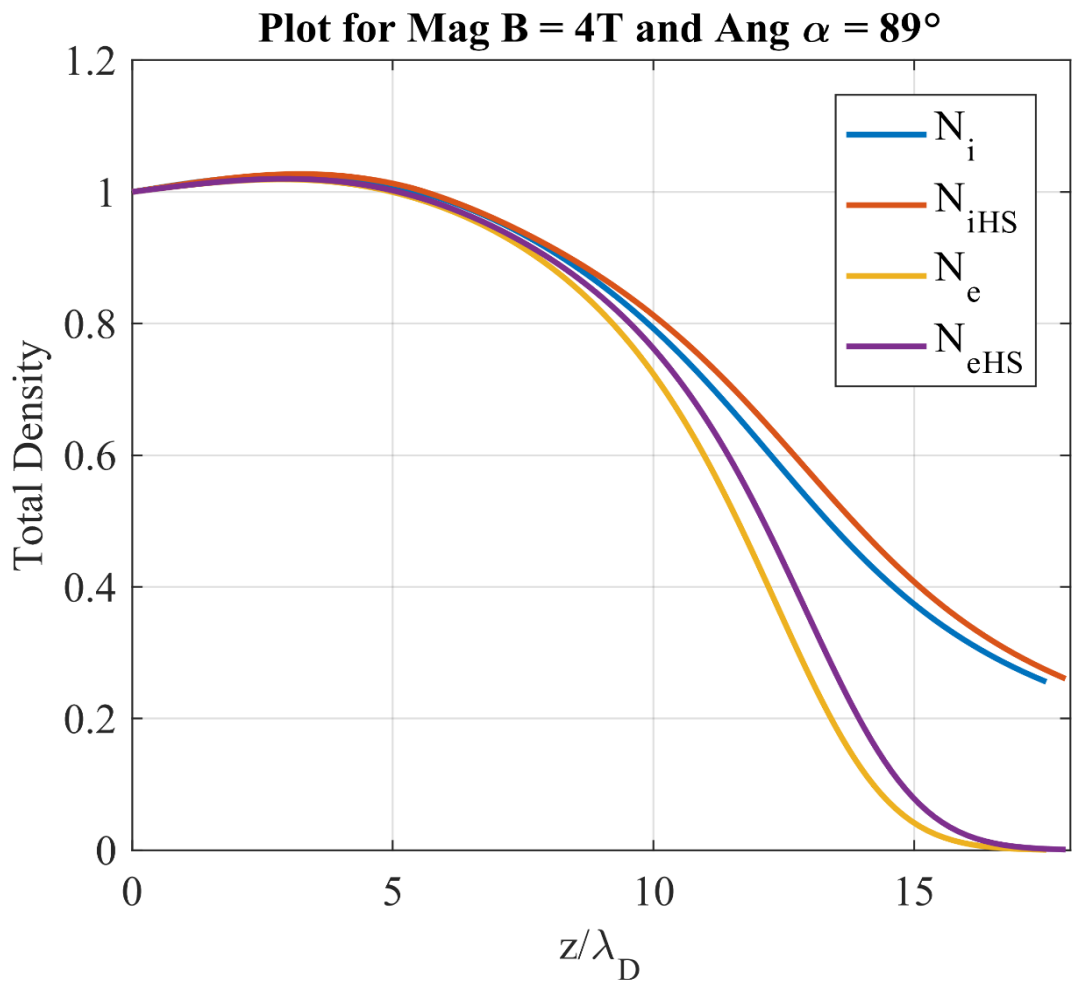

**FIG. 23.** (Color online) The density distributions at a magnetic field strength ($B$) 4T and at a magnetic inclination angle ($\alpha$) 89°.

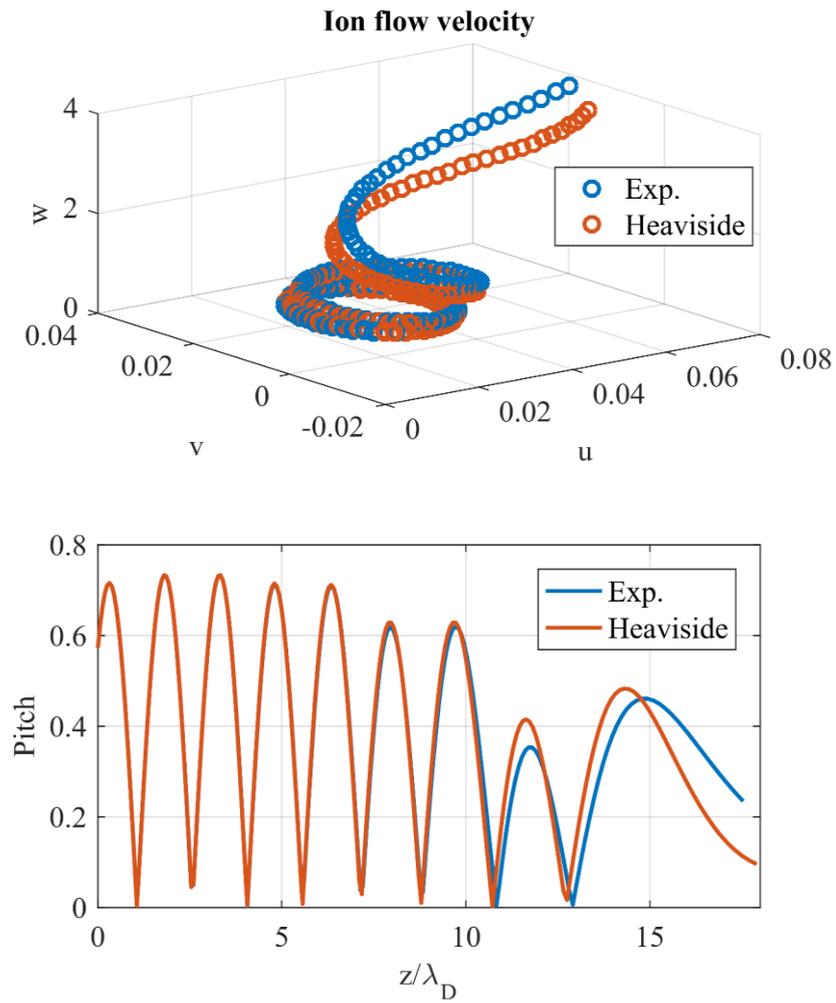

**FIG. 24.** (Color online) The ion flow velocities at a magnetic field strength ($B$) 4T and at a magnetic inclination angle ($\alpha$) 89°.